\documentclass[reprint,nofootinbib, amsmath,amssymb,aps,onecolumn]{revtex4-2}
\usepackage{amssymb,amsmath,amsfonts,amsbsy,graphicx}
\usepackage{color}
\usepackage{psfrag}
\usepackage{stackrel}

\usepackage{csquotes}

\newcommand\be{\begin{equation}}
\newcommand\ee{\end{equation}}
\newcommand\ba{\begin{eqnarray}}
\newcommand\ea{\end{eqnarray}}\newcommand\eq{\begin{equation}}           
\newcommand\en{\end{equation}}

 \newcount\colveccount
\newcommand*\colvec[1]{
        \global\colveccount#1
        \begin{pmatrix}
        \colvecnext
}
\def\colvecnext#1{
        #1
        \global\advance\colveccount-1
        \ifnum\colveccount>0
                \\
                \expandafter\colvecnext
        \else
                \end{pmatrix}
        \fi
}
\input epsf

\usepackage{ulem}
\usepackage{amssymb}
\usepackage{amsmath}
\usepackage{bm}
\usepackage{graphicx}
\usepackage{color}
\usepackage{subfig}
\usepackage{caption}

\def\gsim{\;\rlap{\lower 2.5pt
 \hbox{$\sim$}}\raise 1.5pt\hbox{$>$}\;}
\def\lsim{\;\rlap{\lower 2.5pt
 \hbox{$\sim$}}\raise 1.5pt\hbox{$<$}\;}

\usepackage{graphicx}
\usepackage{url}
\usepackage{amssymb,amsmath}
\usepackage{amsbsy}

\usepackage{graphicx}
\usepackage{url}
\usepackage{amssymb,amsmath}
\usepackage{amsbsy}

\begin{document}
\title{
  Primordial black hole dark matter in the presence of p-wave WIMP annihilation
}
\author{Kenji Kadota$^{1,2,3}$ and Hiroyuki Tashiro$^{4}$ \\
{\small  $^1$School of Fundamental Physics and Mathematical Sciences,
  Hangzhou Institute for Advanced Study,\\
  University of Chinese Academy of Sciences (HIAS-UCAS), Hangzhou 310024, China\\
   $^2$International Centre for Theoretical Physics Asia-Pacific (ICTP-AP), Beijing/Hangzhou, China\\
  $^3$Kobayashi-Maskawa Institute for the Origin of Particles and the Universe, Nagoya University, Nagoya 464-8602, Japan\\
   $^4$Department of Physics and Astrophysics, Nagoya University, Nagoya 464-8602, Japan
}
}

\begin{abstract}
  We study the allowed primordial black hole (PBH) dark matter abundance in the mixed dark matter scenarios consisting of PBHs and self-annihilating weakly interacting massive particles (WIMPs) with a velocity dependent annihilation cross section. We first briefly illustrate how the WIMP dark matter halo profile changes for the velocity suppressed p-wave annihilation scenarios, compared with the familiar s-wave annihilation scenarios, and then discuss the PBH mass dependent upper bound on the allowed PBH dark matter abundance.

  The WIMPs can accrete onto a PBH to form an ultracompact minihalo with a spiky density profile. Such a spike is moderated in the central region of a halo because the WIMPs are annihilated away and this moderation is less effective for a smaller annihilation cross section. The WIMP core density becomes larger while the core radius becomes smaller for a velocity suppressed p-wave annihilation cross section than those for the s-wave annihilation scenarios. 
  The annihilation cross section is dependent on the velocity which varies across the halo, and, in addition to the change of the WIMP density profile, another interesting feature is the PBH mass dependent bound on PBH dark matter abundance. This is in stark contrast to the s-wave annihilation scenarios where the PBH abundance bound is independent of the PBH mass. The allowed PBH dark matter fraction (with respect to the total dark matter abundance) is of order $f_{PBH}\lesssim {\cal O}(10^{-7})(M_{\odot}/M_{PBH})^{(-6+2\gamma_{sp})/(3\gamma_{sp}+3)}$ for the thermal relic p-wave dark matter with the mass $100$ GeV where $\gamma_{sp}$ is the slope index of the spike profile, to be compared with $f_{PBH}\lesssim {\cal O}(10^{-9})$ for the corresponding thermal relic s-wave dark matter scenarios.
\end{abstract}

\maketitle   

\setcounter{footnote}{0} 
\setcounter{page}{1}\setcounter{section}{0} \setcounter{subsection}{0}
\setcounter{subsubsection}{0}


\section{Introduction}
It has been pointed out that the PBH dark matter (DM) and annihilating WIMP DM cannot co-exist \cite{Lacki:2010zf,Tashiro:2021xnj,Yang:2020zcu,Scott:2009tu,Gondolo:1999ef,Lacki:2010zf,Boucenna:2017ghj,Adamek:2019gns,Eroshenko:2016yve,Carr:2020mqm,Cai:2020fnq,Delos:2018ueo,Kohri:2014lza,Bertone:2019vsk,Ando:2015qda,Hertzberg:2020kpm,Yang:2020zcu,Zhang:2010cj,Afshordi:2003zb,Tashiro:2021kcg,Kadota:2020ahr,Inman:2019wvr,Murgia:2019duy,Oguri:2020ldf,Gong:2017sie,Carr:2018rid,Mena:2019nhm}. One reason is that WIMP DM can accrete to a PBH, and the resultant WIMP density enhancement around a PBH (dressed PBH) can lead to an unacceptably large annihilation rate which is proportional to DM density squared.
Such an energy injection into the plasma from the efficient DM annihilation can affect the thermal and ionization history of the Universe.
The consequent stringent bounds $f_{PBH}\lesssim {\cal O}(10^{-10}) \sim{\cal O}(10^{-8})(f_{PBH}\equiv \Omega_{PBH}/\Omega_{DM}$ with $\Omega_{DM}=\Omega_{PBH}+\Omega_{\chi}$) for the WIMP mass range $m_{\chi}\sim  {\cal O}(10)\sim{\cal O}(10^{3})$ GeV were obtained from the CMB (in particular from the CMB polarization) and the 21cm observables \cite{Lacki:2010zf,Tashiro:2021xnj,Yang:2020zcu}. The comparable bounds can also be obtained from the lack of excess signals in the gamma ray observations \cite{Scott:2009tu,Gondolo:1999ef,Lacki:2010zf,Boucenna:2017ghj,Adamek:2019gns,Eroshenko:2016yve,Carr:2020mqm,Cai:2020fnq,Delos:2018ueo,Kohri:2014lza,Bertone:2019vsk,Ando:2015qda,Hertzberg:2020kpm,Yang:2020zcu,Zhang:2010cj}.
  Another independent and complimentary reason for the incompatibility of PBH and WIMP DM can come from the enhancement of small scale structure formation in the presence of PBH entropy fluctuations \cite{Afshordi:2003zb,Tashiro:2021kcg,Kadota:2020ahr,Inman:2019wvr,Murgia:2019duy,Oguri:2020ldf,Gong:2017sie,Carr:2018rid,Mena:2019nhm}. Such bounds are independent of the WIMP profile around a PBH and hence of particular interest for the light (sub-GeV) WIMPs and/or light (sub-solar mass) PBHs for which the ultracompact minihalo (UCMH) profile around a PBH can be affected by the non-negligible WIMP kinetic energy at the halo formation epoch. Randomly distributed PBHs can lead to the Poisson noise in the matter density and they can dominate the conventional adiabatic perturbations at small scales. Such PBH-sourced fluctuations enhance the small scale structure formation, which results in abundant small halos which are dense because they are formed early (at the redshift above 100 well before the typical halo formation epoch ($z\lesssim 20$) in the conventional $\Lambda$CDM without PBHs). The bounds from CMB and 21cm as well as from gamma ray observables due to the annihilation from these abundant small halos, even though they are not as dense/compact as the UCMHs around PBHs, have also been studied to demonstrate the incompatibility of the mixed DM scenarios consisting of PBHs and WIMPs \cite{Kadota:2020ahr,Tashiro:2021kcg}.

  The DM annihilation in the mixed PBH-WIMP DM scenarios so far has been mainly discussed for the velocity independent annihilation cross section.
  If the annihilation cross section is velocity suppressed however the bounds on the allowed PBH fraction can be relaxed. The goal of this paper is to study how much the p-wave dominated WIMP annihilation can relax the bounds on the allowed PBH abundance. We illustrate our findings using the gamma ray bounds due to DM annihilation from the UCMHs around PBHs. The quantitative effects of annihilation suppression  may not be so obvious because, even though the annihilation rate can be velocity suppressed, the DM core density can be enhanced in the velocity suppressed annihilation scenarios. The DM spike profile is saturated in the inner region of spike because the WIMP DM is annihilated away, and this maximum WIMP density is inversely proportional to the annihilation rate. The DM annihilation flux which is proportional to DM density squared hence can benefit from such an enhanced DM core density. A characteristic feature for p-wave scenarios is a non-trivial dependence of $f_{PBH}$ bound on the PBH mass, to be compared with the PBH mass independent bound of $f_{PBH}$ for s-wave annihilation scenarios. Throughout the paper, we assume $f_{PBH}\ll 1$ (equivalently the WIMP fraction $f_{\chi}= 1-f_{PBH} \sim 1$) in our discussions unless stated otherwise which will be justified in our quantitative analysis.

  It is common to expand the thermally averaged cross section times velocity as
$ \langle \sigma v \rangle = a+ b x^{-1} + ...$, where $a,b$ are constant and represent respectively s-wave and p-wave contributions. We for concreteness focus on the p-wave annihilation scenarios in the main body of the paper. The generalization to other partial wave annihilation scenarios is straightforward and is briefly discussed in the discussion section. The corresponding thermal relic abundance reads 
  \ba
  \left( \frac{\Omega_{DM}h^2}{0.1} \right) \sim 0.3 \left.  \left( \frac{x}{\sqrt{g_*}} \right) \left( \frac{3\times 10^{-26} cm^3/s}{a+ \frac{b}{2}x^{-1}} \right) \right\vert_{fo}
\ea
where $g_*$ is the relativistic degrees of freedom, $x=m_{\chi}/T$ and \enquote*{$fo$} indicates the right hand side is meant to be evaluated at the DM freeze out \cite{kolb1990,Jungman:1995df} \footnote{Another common convention is to expand the cross section in power of $v^2$ before taking the thermal average $\langle  \sigma v \rangle =\langle a+bv^2+... \rangle= a+ 3 b/2 x^{-1}+...$ \cite{Gondolo:1990dk}.}. The commonly quoted reference values, typically for the neutralino DM which can give the desired DM relic abundance, are $\langle \sigma v \rangle^s=3 \times 10^{-26} cm^3/s$ for the s-wave annihilation dominated scenarios and $x_{fo}\sim 20$ corresponding to the DM velocity at freeze out $v_{fo}\sim 0.3$.
We parameterize the cross section for the p-wave dominated scenarios as $\langle \sigma v \rangle = \langle \sigma v \rangle^p_0 (v/v_{0})^2$. $\langle \sigma v \rangle_{0}^p$ and $v_{0}$ are the reference constant values and we use for concreteness the values close to those at the freeze out to match the desired DM relic abundance $\langle \sigma v \rangle_0^p=6\times 10^{-26} cm^3/s$ and $v_{0}=0.3$ in our quantitative discussions unless stated otherwise. 
The goal of our paper is to illustrate the quantitative difference of the allowed PBH parameter range when the p-wave annihilation is dominant compared with the $s$-wave annihilation dominated scenarios commonly discussed in the literature. The $p$-wave contribution can be dominant, for instance, when Majorana DM pair annihilation into a CP even scalar pair through s-wave channel is prohibited by the CP-conservation (see for instance Ref. \cite{Kumar:2013iva} for the summary of interaction operators for which $s$-wave contribution is absent or suppressed). We simply assume in our analysis that the $p$-wave annihilation contribution is dominant without specifying the concrete models to keep our analysis as generic as possible.

We first outline the DM density profile of an ultracompact minihalo around a PBH and the corresponding velocity dispersion in Section \ref{modelsection}. We then present the results on the allowed PBH fraction in the presence of WIMP annihilation in Section \ref{boundsec}, followed by the discussion/conclusion.

\section{WIMP dark matter halo profile around a PBH}
\label{modelsection}
\begin{figure}[!htbp]

     \begin{tabular}{c}

                        \includegraphics[width=0.5\textwidth]{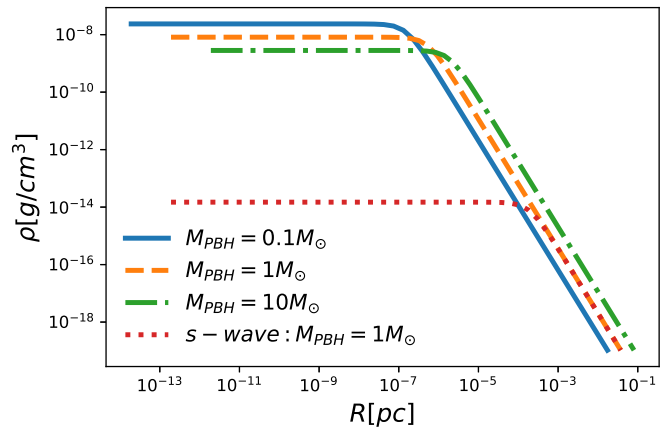} 
    \end{tabular}
       

    
     \caption{The WIMP density profile around a PBH for the thermal relic p-wave dark matter. The WIMP mass $m_{\chi}$ is set to 100 GeV. Each plot covers the radius from $R_{min}=4GM_{PBH}$ to the turn around radius at matter radiation equality $R_{ta}(z_{eq})$. The corresponding density for the thermal relic s-wave dark matter with $M_{PBH}=1M_{\odot}$ is also shown for comparison.}
   \label{rhopands}
\end{figure}

\begin{figure}[!htbp]

     \begin{tabular}{c}
                               \includegraphics[width=0.5\textwidth]{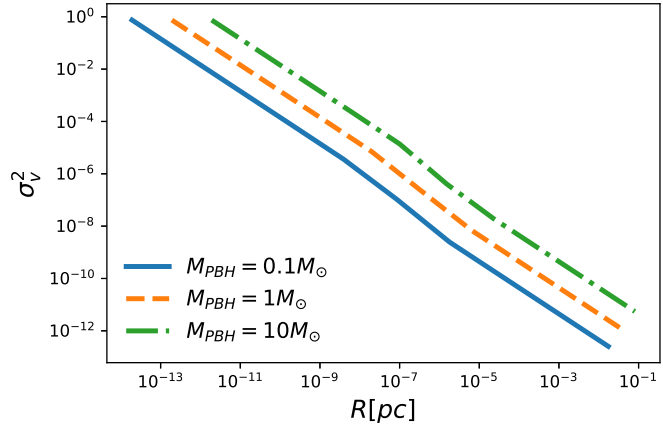} 
    \end{tabular}
       
    
     \caption{
       The WIMP velocity dispersion around a PBH. $m_{\chi}$ is set to 100 GeV. Each plot covers the radius from $R_{min}=4GM_{PBH}$ to the turn around radius at matter radiation equality $R_{ta}(z_{eq})$.}
   \label{velprofile}
\end{figure}

 We first outline the modeling of a dressed PBH (PBH 'dressed' by WIMP DM) due to the WIMP DM accretion to a PBH. 
 Following the formation of a PBH in the radiation dominated epoch, the DM particle orbits can be bound to a PBH to form the UCMH. The UCMH size can be estimated by the turn around radius where the gravitational influence of PBH decouples the DM from the background Hubble flow \cite{Adamek:2019gns}
   \ba
   r_{ta}\approx (R_S  t_{ta}^2)^{1/3}
   \label{eqrta}
 \ea
 where the Schwarzschild radius $R_S=2GM_{PBH}$ and $t_{ta}$ is the time when the shell turns around at $r_{ta}$. One can analytically estimate the DM density around a PBH by assuming the DM density of each mass shell matches the background density at the corresponding turn-around epoch. The DM density profile during the radiation domination epoch then becomes $\rho_{DM}(r)= \rho(r(t_{ta})) \sim  (\rho_{eq}/2) (t_{ta}/t_{eq})^{-3/2} \propto r^{-9/4}$ using Eq. \ref{eqrta}.
The spike slope (in this simple example it is $-9/4$) in the inner halo region $r < r_{ta}(z_{eq})$ (turn around scale at matter radiation equality) is numerically shown to be kept even when DM halo grows by the secondary infall mechanism during the matter domination epoch in the region $r > r_{ta}(z_{eq})$ where the halo mass, rather than PBH mass, influence on the halo formation can be dominant \cite{Adamek:2019gns,Carr:2020mqm}. Such a DM density of a halo formed during the matter domination epoch does not exceed the spike density at $r=r_{ta}(z_{eq})$ and, considering that the DM annihilation is proportional to the DM density squared, a big density in the spike region can well exceed the annihilation flux from the outer halo region (which can follow, for instance, the conventional NFW profile \cite{Eroshenko:2016yve,Adamek:2019gns}). There has been pointed out that the inner most central region can be shallower than the outer spike slope $\gamma_{sp}$, but such a change in the profile slope does not significantly affect the gamma ray signals in the parameter range of our interest because the central part of the DM halo is annihilated away \cite{Eroshenko:2019pxt, Carr:2020mqm}. The flattening of the inner central profile due to DM annihilation can be taken account of by considering the maximum DM density
 \ba
 \rho_{core}=\frac{m_{\chi}}{\langle \sigma v \rangle t_{bh}}
 \ea
 where $t_{bh}$ is the age of PBH \cite{Ullio:2002pj,Bringmann:2011ut,Berezinsky:2005py}. We therefore model the UCMH profile around a PBH as
  \cite{Gondolo:1999ef,Tashiro:2021xnj,Shelton:2015aqa, Yang:2020zcu,Fields:2014pia,Shelton:2015aqa,Shapiro:2016ypb,Johnson:2019hsm,Adamek:2019gns,Boucenna:2017ghj,Boudaud:2021irr,Carr:2020mqm,Eroshenko:2016yve,Ullio:2001fb}
 \ba
\rho(r) = 
 \begin{cases}
  0     \quad &{\rm for}~ r < 4GM_{PBH}  \\
   \frac{\rho_{sp}(r) \rho_{in}(r)}
       {\rho_{sp}(r)+\rho_{in}(r)}
  \quad &{\rm for}~ 4GM_{PBH} \leq r < r_{\rm ta}(t_{\rm eq}) . \\
 \end{cases}
 \label{rho1b}
 \ea
  where
 \ba
 \rho_{in}(r)=\rho_{pl}(t) (r_{cut}/r)^{\gamma_{in}}
 \\
  \rho_{sp}(r) =\rho_{pl}(t)(r_{cut}/r)^{\gamma_{sp}}
  \ea
The DM particle is captured by PBH and hence DM density vanishes for $r<2R_{S}=4GM_{PBH}$ ($R_S$ is the Schwarzschild radius). This radius $4GM_{PBH}$ corresponds to the radius of marginally bound DM orbits and the minimum periastron of parabolic orbits around a Schwarzschild black hole \cite{Vasiliev:2007vh,Sadeghian:2013laa}. We use the slope of spike profile $\gamma_{sp}=9/4$ as a fiducial value in our quantitative discussions unless stated otherwise \cite{Adamek:2019gns,Serpico:2020ehh,Gondolo:1999ef,Boudaud:2021irr,Eroshenko:2016yve, Boucenna:2017ghj} \footnote{Some literature discusses a value even up to $\gamma_{sp} \sim 2.75$, but our qualitative discussion and conclusion are not affected by such a change in the value of $\gamma_{sp}$ \cite{Gondolo:1999ef,Johnson:2019hsm,Fields:2014pia}.}. We also use for concreteness the flat annihilation plateau $\rho_{in}(r)=\rho_{core}$, namely $\rho_{pl}(t)=\rho_{core},\gamma_{in}=0$ in our discussions unless stated otherwise and the spike profile given by Eq. \ref{rho1b} can be obtained by integrating $\dot{\rho}=-\sigma v {\rho^2}/{m_{\chi}}$. Note the amplitude of the annihilation plateau decreases over time as expected. $\rho_{in}$ represents the inner spike profile and $\rho_{sp}$ does the outer spike profile, and $r_{cut}$ represents the radius where $\rho_{in}(r_{cut})=\rho_{sp}(r_{cut})$. It also has been pointed out that the inner region $r\lesssim r_{cut}$ can possibly deviate from the flat plateau (the flat plateau is valid only for the circular orbit of DM around a PBH for s-wave annihilation scenarios) \cite{Vasiliev:2007vh}). For the p-wave annihilation scenarios, it is not flat even for the circular orbit because the annihilation rate is dependent on a radius. The cases for the annihilation plateau profile $\rho_{in}\propto r^{-\gamma_{in}}$ with $\gamma_{in}\sim 0.5$ for the s-wave and $\gamma_{in}\sim 0.3$ for the p-wave dominated scenarios have been discussed in the literature \cite{Shapiro:2016ypb, Vasiliev:2007vh}. As pointed out in \cite{Vasiliev:2007vh}, however, the non-zero slope of annihilation plateau does not significantly affect the following discussions on the observation constraints, unless the experiment can resolve such small regions, because the occupied volume is small in the inner part of the spike. We found, as shown in the next section, that the order of magnitude for the upper bounds on the PBH fraction does not change among $\gamma_{in}=0, 0.3, 0.5$ and hence does not affect our following discussions.
In fact, the dominant annihilation contribution comes from the region around $r\sim r_{cut}$ (see Fig. \ref{rhopands} for $r_{cut}$ which represents the core size and the corresponding velocity in Fig. \ref{velprofile}). We aim to obtain the bounds on the allowed PBH DM fraction $f_{PBH}$ by considering the enhanced annihilation flux due to such a spike profile.

 
 In addition to the DM density profile, one also needs to specify the DM velocity to estimate the velocity dependent annihilation cross section to illustrate the comparison between the s-wave and the p-wave annihilation scenarios. We consider the velocity dispersion $\sigma_v^2$ to estimate the effect of this velocity suppression in the thermally averaged cross section $\langle \sigma v \rangle$ (which is motivated from the Maxwellian velocity distribution $f_{MB}(v)$ for which the thermally averaged p-wave cross section $\langle \sigma v \rangle_p \propto \int dv v^2 f_{MB}(v) =\sigma_v^2$). The relevant velocity dispersion can be modeled by matching a piece-wise solution of Jeans equation which relates velocity dispersion and mass density in dynamical equilibrium \cite{2008gady.book.....B}
\ba
 \sigma_v^2(r) = 
 \begin{cases}
  \frac{GM_{PBH}}{r}
  \frac{3}{1+\gamma_{in}}
  \left[
    1+\frac{r}{r_{cut}}
    \left(
\frac{\gamma_{in}-\gamma_{sp}}{1+\gamma_{sp}}
    \right)
    \right]
     \quad &{\rm for}~ 4GM_{PBH}< r < r_{\rm cut}(z)  \\
     \frac{GM_{PBH}}{r}
\frac{3}{1+\gamma_{sp}}   \quad &{\rm for}~r_{\rm cut}(z) \leq r < r_{\rm ta}(z_{\rm eq}) \\
  \end{cases}
 \ea
 where we assumed the isotropic velocity dispersion profile in solving the spherical Jeans equation \cite{Fields:2014pia,Shapiro:2016ypb}. The DM density and corresponding velocity profiles are shown respectively in Fig. \ref{rhopands} and  Fig. \ref{velprofile}.
 The radius range for each curve shown in these figures is from the turn around radius at matter radiation equality $R_{ta}(z_{eq})\sim 0.04 (M_{PBH}/M_{\odot})^{1/3} $ pc down to $R_{min}=4GM_{PBH}\sim 2\times 10^{-13} (M_{PBH}/M_{\odot})$pc. 
Even though the annihilation flux becomes smaller due to the velocity suppressed annihilation cross section for p-wave scenarios compared with s-wave scenarios, this suppression is partially compensated by a larger core density. The amplitude of core density is bigger for the p-wave case because DM is less annihilated away. The radial size of core $r_{cut}$ accordingly becomes smaller for p-wave compared with s-wave scenarios. As we discuss later, these changes lead to the PBH mass dependent bounds on the allowed $f_{PBH}$ in contrast to s-wave scenarios where the upper bound on $f_{PBH}$ is independent of PBH mass. We also mention that the DM velocity is non-relativistic around $r\sim r_{cut}$ where the dominant annihilation signal comes from, which justifies our not including the relativistic corrections in our analysis and we leave the studies including relativistic corrections for future work \cite{Eroshenko:2019pxt,Sadeghian:2013laa}.

 \section{Allowed PBH dark matter abundance}
 \label{boundsec}
\begin{figure}[!htbp]

     \begin{tabular}{c}
                               \includegraphics[width=0.5\textwidth]{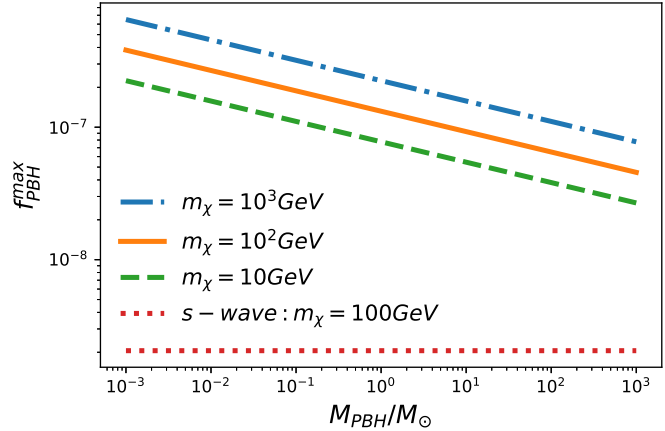} 
    \end{tabular}
       
    
     \caption{
      The upper bound on PBH DM fraction $f_{PBH}(\equiv \Omega_{PBH}/\Omega_{DM}$) in the presence of thermal relic p-wave dark matter. The WIMP mass and density profile parameters are set to $m_{\chi}=100GeV, \gamma_{in}=0, \gamma_{sp}=9/4$. The corresponding bound for thermal relic s-wave dark matter (with $m_{\chi}=100 GeV, \gamma_{in}=0, \gamma_{sp}=9/4$) is also shown for comparison.}
   \label{fpbhmaxfigdec2}
\end{figure}

 As a concrete example of possible constraints on the allowed PBH DM fraction, we apply the diffuse gamma ray background bounds on the mixed PBH-WIMP scenarios with a velocity dependent WIMP annihilation cross section. The signal from WIMP annihilation within unresolved regions morphologically resembles that of the decaying (or shining) dressed PBH DM, and the analysis is similar to that of the decaying DM. More specifically, the bounds on the PBH fraction can be obtained by equating the total dressed PBH decay rate to the total WIMP decay rate \cite{Lacki:2010zf,Boucenna:2017ghj,Adamek:2019gns}
 \ba
\frac{f_{PBH} \Gamma_{PBH}}{M_{PBH}}=\frac{\Gamma_{\chi}}{m_{\chi}}
\label{fpbhconvert}
\ea
where the dressed PBH can be morphologically interpreted as the decaying DM with a mass $M_{PBH}$ and the decay rate 
\ba
\Gamma_{PBH}=\frac{1}{m_{\chi}^2} \int dR 4\pi R^2 \rho^2 \langle\sigma v(R) \rangle
\label{pbhdecayrate}
\ea
The integration is over the volume of interest and we consider the annihilation flux from the region covering $R=4GM_{PBH}$ to $R=r_{ta}(t_{eq})$. Following Ref. \cite{Adamek:2019gns} for an easier comparison, we apply the WIMP decay bounds $\Gamma_{\chi}^{-1}\gtrsim 10^{28}$s which was obtained in Ref. \cite{Ando:2015qda} as a bound on the decaying DM scenarios (with the $b\bar{b}$ decay channel) for the Fermi sensitive mass range $10$ GeV $\lesssim m_{\chi}\lesssim 10^4$GeV.\footnote{See Ref. \cite{Ando:2015qda} for other decay channels besides $b\bar{b}$ and the corresponding bounds on $f_{PBH}$ can be obtained by the straightforward scaling of our results.} The DM density $\rho$ depends on the annihilation cross section $\langle\sigma v \rangle $ because $r_{cut}$ depends on it and so does the amplitude of the annihilation plateau. The velocity dependent annihilation cross section hence can lead to the non-trivial dependence of allowed $f_{PBH}$ on the model parameters. We here pay a particular attention to the dependence on $M_{PBH}$. Fig. \ref{fpbhmaxfigdec2} shows the allowed upper bound on PBH DM fraction. Note this bound is independent of $M_{PBH}$ for s-wave annihilation scenarios, because $\Gamma_{PBH}\propto M_{PBH}$ (the decay is bigger for a heavier PBH mass) cancels out the $M_{PBH}$ factor in the denominator in Eq. \ref{fpbhconvert} (the number density of PBH becomes smaller for a heavier PBH mass).
For p-wave dominated annihilation scenarios, there is a non-trivial $M_{PBH}$ dependence in the allowed $f_{PBH}^{max} \propto M_{PBH}^{\alpha}$ with $\alpha=\frac{-6+2\gamma_{sp}}{3(\gamma_{sp}+1)}$. The choice of $\gamma_{sp}$ values vary among the literature and some use the value even up to $\gamma_{sp}\sim 2.75$ \cite{Fields:2014pia,Gondolo:1999ef}. We used $\gamma_{sp}=9/4$ in Fig. \ref{fpbhmaxfigdec2} as analytically argued in Sec \ref{modelsection}.
 \begin{figure}[!htbp]
  \begin{center}
     \begin{tabular}{cc}
       \includegraphics[width=0.48\textwidth]{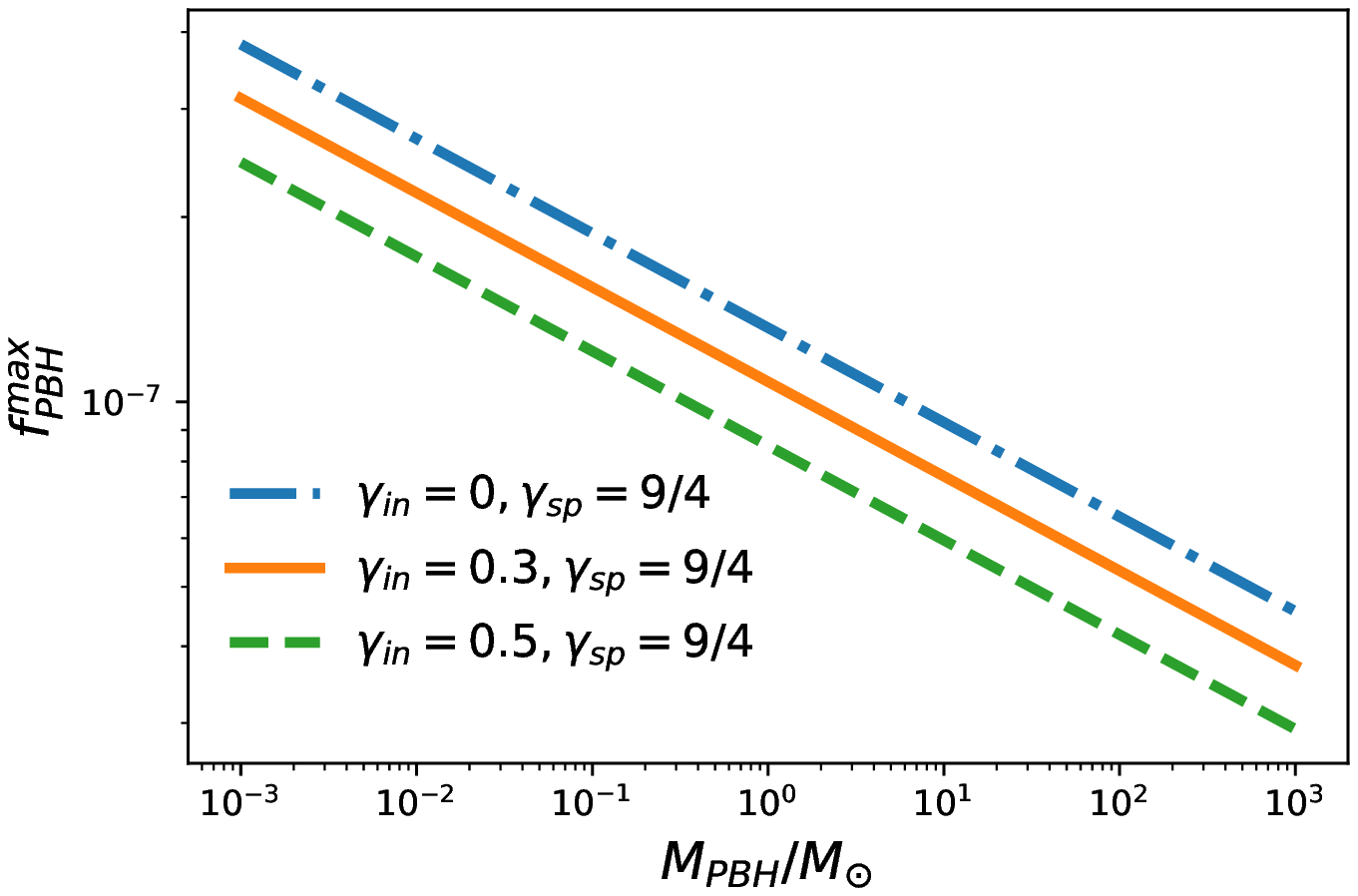} &
       \includegraphics[width=0.48\textwidth]{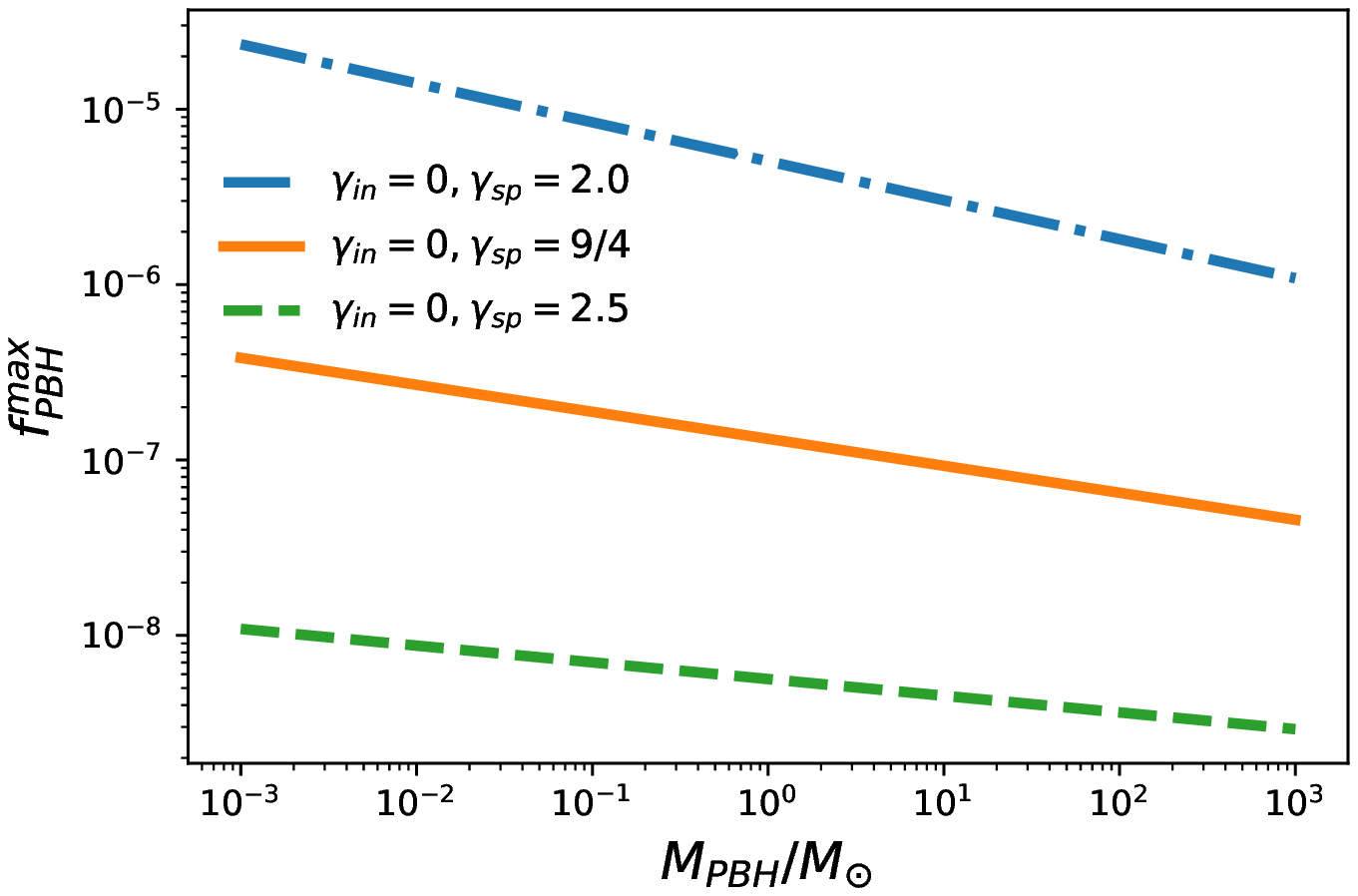}
         \\
    \end{tabular}
  \end{center}
     \caption{
      The dependence of $f_{PBH}$ upper bound on the halo profile in the presence of thermal relic p-wave dark matter. The WIMP mass is set to 100 GeV.}
   \label{fmaxprofilefig}
\end{figure}
\begin{figure}[!htbp]

     \begin{tabular}{c}
                               \includegraphics[width=0.5\textwidth]{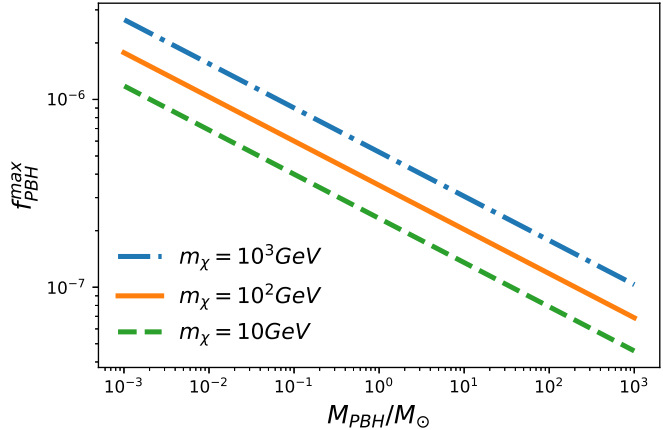} 
    \end{tabular}
       
    
     \caption{
      The upper bound on the PBH DM fraction $f_{PBH}$ in the presence of thermal relic d-wave dark matter. The halo profile parameters are set to $\gamma_{in}=0,\gamma_{sp}=9/4$.}
   \label{dwave}
\end{figure}
There are a few competing effects which can affect the scaling relation between $f_{PBH}^{max}$ and $M_{PBH}$ illustrated in Fig. \ref{fpbhmaxfigdec2}. One is the suppression of the annihilation rate $\langle \sigma v \rangle $ factor in Eq. \ref{pbhdecayrate} which can reduce the annihilation flux. The PBH mass also affects the density $\rho^2$ factor in Eq. \ref{pbhdecayrate}. The core amplitude $\rho_{core}$ in the inner spike region is inversely proportional to $\langle \sigma v \rangle $ and the suppression of $\langle \sigma v \rangle $ can help enhance the density amplitude in the core. The volume of inner core region however is small as can be seen in Fig. \ref{rhopands}, and the most of the annihilation signal comes from the region around $r\sim r_{cut}$. $r_{cut}$ increases for a smaller core amplitude and hence the volume around $r_{cut}$ is bigger for a bigger $\langle \sigma v \rangle $. Note $r_{cut}$ represents the radius where the annihilation rate equals the Hubble rate ($\sim 1/t_{bh}$) and the annihilation rate at $r=r_{cut}$ is common for both $p$-wave and $s$-wave annihilation scenarios. The $r_{cut}$ dependence on $M_{PBH}$ plays a key role in a non-trivial $f_{PBH}$-$M_{PBH}$ scaling relationship.
  Fig. \ref{fpbhmaxfigdec2} also shows the $f_{PBH}^{max}$ dependence on $m_{\chi}$. The upper bound on $f_{PBH}$ becomes weaker for a bigger $m_{\chi}$. This behavior arises because the WIMP number density decrease for a bigger $m_{\chi}$ which can cancel the effect of $\rho_{core}\propto m_{\chi}$ increase for a bigger $m_{\chi}$, as can be seen in Eqs. \ref{fpbhconvert},\ref{pbhdecayrate}. Another effect of the change in $m_{\chi}$ is also reflected in the change of $r_{cut}$ which decreases for a bigger $m_{\chi}$. 
 We also mention the dependence of the $f_{PBH}$ bounds on the slope of halo profile. While the slope of the inner spike region $\gamma_{in}=0$ can be justified for the DM circular orbit for s-wave scenarios, $\gamma_{in}\sim 0.5$ has been discussed for more general orbits \cite{Vasiliev:2007vh,Shapiro:2016ypb}. For the velocity dependent annihilation scenarios, on the other hand, the inner core plateau would not be flat even for the circular orbit because the annihilation rate is dependent on the radius and $\gamma_{in}\sim 0.3$ has been discussed for p-wave annihilation scenarios \cite{Shapiro:2016ypb,Johnson:2019hsm}. Fig. \ref{fmaxprofilefig} shows how the upper bounds on $f_{PBH}$ varies by a different choice of $\gamma_{in}$. The volume inside the annihilation plateau region is small and varying $\gamma_{in}$ would not significantly affect our discussions. We however mention that our bounds can be more sensitive on the value of $\gamma_{sp}$  as shown in Fig. \ref{fmaxprofilefig}. This behavior arises because a steeper profile (a bigger $\gamma_{sp}$) leads to a bigger core radius $r_{cut}$. Even though the bounds on the allowed PBH DM fraction can be relaxed compared with s-wave scenarios, we conclude that the incompatibility of PBH and WIMP DM still holds even if one considers the velocity dependent WIMP annihilation cross sections.
 
 Before concluding our discussions, let us briefly mention the straightforward extension of our discussions to other partial wave dominated annihilation scenarios. For $\langle \sigma v \rangle
\propto
(v/v_0)^{2n}
\propto
\left({M_{PBH}}/{r}
\right)^n
$, the dependence of the bounds scales as $f^{max}_{PBH}\propto M_{PBH}^{\alpha}$ with $\alpha=\frac{n(-6+2\gamma_{sp})}{3(\gamma_{sp}+n)}$. We show the results for d-wave ($n=2$) as an example in Fig. \ref{dwave}. We used the normalization constant  $\langle \sigma v \rangle^s =\langle \sigma v \rangle^d_0/3=3\times 10^{-26}cm^3/s$
in parameterizing the cross section $\langle \sigma v\rangle =\langle \sigma v \rangle_0 (v/v_0)^{2n}$, which is motivated from the parameterization of  $\langle \sigma v \rangle = \sum_{n=0}^{\infty} c_n x^{-n} $ leading to 
\ba
\left( \frac{\Omega_{DM}h^2}{0.1} \right) \sim 0.3 \left.  \left( \frac{x}{\sqrt{g_*}} \right) \left( \frac{3\times 10^{-26} cm^3/s}{  \sum_{n=0}^{\infty}
  \frac{c_n}{n+1}
  x^{-n}  } \right) \right\vert_{fo}
\ea 
The WIMP core density in the inner region of spike can become larger for a velocity suppressed annihilation cross section, compared with that of velocity independent cross section scenarios. Nevertheless, the inner region volume is small and overwhelmed by the larger radius region with a velocity suppressed annihilation cross section, leading to a weaker bound on the allowed PBH fraction compared with the velocity independent annihilation cross section scenarios. Even though the allowed PBH DM fraction is not as tight as the velocity independent annihilation scenarios, the PBH DM fraction should be still negligible compared with WIMP DM (and vise versa). 
We also mention that, for PBHs with masses above a few tens of solar mass, even the baryonic accretion effects without considering WIMP annihilation can give the stringent bounds from the CMB ($f_{PBH}\ll {\cal O}(10^{-3})$) \cite{Ricotti:2007jk,Mack:2006gz,Ricotti:2009bs,Ricotti:2007au,Poulin:2017bwe,Serpico:2020ehh}, even though the bounds from gas accretion become significantly weak for light PBH masses below a solar mass \footnote{The WIMP annihilation energy scale is of order WIMP mass while the injected energy scale from the baryon accretion is at most of order MeV. The typical weak-scale WIMP annihilation bounds can hence be more stringent than the baryon accretion bounds \cite{Tashiro:2021xnj}.}. 
The validity of our WIMP annihilation bounds for a small PBH mass below the range discussed in our analysis depends on whether the WIMP kinetic energy can be still negligible compared with the WIMP potential energy at the halo formation epoch.
So far we have discussed the WIMP DM which are gravitationally clustered to form a halo, and estimated the WIMP velocity dispersion around a PBH assuming the spherical symmetry and steady-state hydrodynamic equilibrium. In addition to such DM clumped around a PBH, there exists uniformly distributed (unclustered) DM whose velocity can be heavily dependent on the nature of DM kinetic decoupling. This concerns the elastic scattering rate with the cosmic plasma \cite{Loeb:2005pm,Bertschinger:2006nq,Gondolo:2012vh, Profumo:2006bv,Gondolo:2016mrz,Green:2003un,Green:2005fa,Bringmann:2006mu}, rather than a self annihilation rate we have discussed in this paper. 
We in fact assumed that DM halo is formed around a PBH under the influence of gravity due to the PBH mass, but the gravitational influence may well be disturbed if the WIMP's kinetic energy is not negligible compared with its potential energy. Such cases could occur for a sufficiently light PBH mass or light WIMP mass, and the WIMP halo density profile may well be affected by the non-negligible kinetic energy. More specifically, assuming a simple scaling of the DM kinetic decoupling and the mass $T_{KD}\propto m^{5/4}_{\chi}$ typical for the bino-like WIMP, the kinetic energy is less than a percent level of the potential energy for $M_{PBH}\gtrsim 10^{-6} M_{\odot}$ if $m_{\chi}\gtrsim 100$ GeV \cite{Adamek:2019gns,Carr:2020mqm,Kadota:2020ahr}. The detailed numerical simulations for the DM profile when the kinetic energy is non-negligible at the halo formation epoch still have not been performed yet, but the analytical estimation assuming the Boltzmann velocity distribution for WIMPs shows a significantly weak bound on $f_{PBH}$ for such cases \cite{Carr:2020mqm,Boudaud:2021irr,Boucenna:2017ghj}. We leave such scenarios where the WIMP kinetic energy cannot be neglected compared with the potential energy for future work. We also leave the study with the PBH mass distribution beyond the monochromatic mass distribution assumed in our discussions for future work. Other observations besides the gamma ray signals, such as CMB and 21cm \cite{Tashiro:2021xnj}, would be also worth pursuing when the WIMP annihilation cross section is velocity dependent, which we plan to study in the forthcoming paper.
\\
\\
This work was in part supported by Grants-in-Aid for Scientific Research from JSPS (21K03533).
KK thanks Kobayashi-Maskawa institute at Nagoya University for hospitality through JSPS core-to-core program (JPJSCCA20200002) and Grant-in-Aid for Scientific research from the Ministry of
Education, Science, Sports, and Culture (MEXT), Japan (16H06492).

 \bibliography{../kenjireference}



\end{document}